\documentclass[a4paper]{article}
\usepackage{caption}
\usepackage{subcaption}
\usepackage{INTERSPEECH2022}
\usepackage{color}
\usepackage{url}
\usepackage{longtable}
\usepackage[numbers,sort&compress]{natbib}
\usepackage[colorlinks=True]{hyperref}

\title{Streaming End-to-End Multilingual Speech Recognition\\ with Joint Language Identification}
\name{C. Zhang, B. Li, T.N. Sainath, T. Strohman, S. Mavandadi, S. Chang, P. Haghani}
\address{Google LLC, USA}
\email{\{chaoz,boboli,tsainath,strohman,sepand,shuoyiin,parisah\}@google.com}

\begin{document}

\maketitle
\begin{abstract}
Language identification is critical for many downstream tasks in automatic speech recognition (ASR), and is beneficial to integrate into multilingual end-to-end ASR as an additional task. In this paper, we propose to modify the structure of the cascaded-encoder-based recurrent neural network transducer (RNN-T) model by integrating a per-frame language identifier (LID) predictor. RNN-T with cascaded encoders can achieve streaming ASR with low latency using first-pass decoding with no right-context, and achieve lower word error rates (WERs) using second-pass decoding with longer right-context. By leveraging such differences in the right-contexts and a streaming implementation of statistics pooling, the proposed method can achieve accurate streaming LID prediction with little extra test-time cost. Experimental results on a voice search dataset with 9 language locales shows that the proposed method achieves an average of 96.2\% LID prediction accuracy and the same second-pass WER as that obtained by including oracle LID in the input. 
\end{abstract}

\section{Introduction}
With the rapid development of deep learning, ASR performance has significantly improved and has been applied to an increasing number of products used by many people. Although most ASR studies focus on only one language, having mixed languages in an utterance or multiple languages in a dialogue is very common in real-world applications, since approximately 43\% of people in the world are bilingual with a further 17\% being multilingual \cite{ilang2018}. 
Hence it is required in practice to develop truly multilingual ASR systems that can handle a list of languages without knowing the language identity of a test utterance. Furthermore, in downstream tasks, such as code-switching and speech translation, both text transcriptions and their language identifiers (LIDs) are often requested together. It is beneficial to produce them together using a single multitask model built in a fully end-to-end (E2E) way.

Although E2E multilingual ASR \cite{Cho2018,Pratap2020,Li2021} and language identification \cite{Fer2015,Gonzalez2015,Pesan2016,Snyder2018,Wang2019,Wan2019,Titus2020,Chandak2020,Punjabi2020,Tjandra2021} can be studied separately, there exists a large number of previous work on using LID to improve multilingual E2E ASR \cite{Kim2018,Toshniwal2018,Waters2019,Gaur2021,Dalmia2021,Zhou2022,Weiner2012,Dominguez2015,Yang2018,Luo2018,Zeng2019,Shan2019,Li2019,Punjabi2021,Joshi2021,Watanabe2017,Seki2018,Hou2020,Zhang2021,Hu2019,Yan2021}. One body of work shows that using oracle LID information helps multilingual models \cite{Kim2018,Toshniwal2018,Waters2019,Gaur2021,Dalmia2021,Zhou2022}. Our work differs from this as we predict the LIDs instead of using the oracle ones. Another body of work looks at techniques to predict LID and use the predictions in the ASR system or downstream tasks \cite{Weiner2012,Dominguez2015,Yang2018,Luo2018,Zeng2019,Shan2019,Li2019,Punjabi2021,Joshi2021}. Much of this work focuses on LID predictions in a non-streaming system \cite{Yang2018,Luo2018,Zeng2019,Shan2019,Watanabe2017,Seki2018,Hou2020}, which does not fit into our streaming ASR setup that is important due to production constraints. 
Our work also differs from existing streaming 
work \cite{Li2019,Punjabi2021,Joshi2021}, as we use an RNN-T cascaded encoder model with a 0.9-second delay in the second-pass, making it suitable for accurate LID predictions. Furthermore, our LID predictor is lightweight as it uses non-parametric streaming statistics pooling and increases parameters by only 0.5\%. Finally, compared to previous works, our experiments were conducted on a larger-scale of 9 language locales \cite{Li2019,Punjabi2021,Joshi2021}. 

In this paper, we work towards building a truly multilingual ASR system without using any oracle LID information during the test, which requires to have not only similar speech recognition accuracy as the systems with oracle LID inputs, but also the additional ability to predict LIDs in a streaming way. To achieve these goals, we propose to include a LID predictor into an E2E multilingual ASR based on RNN-T with cascaded encoders \cite{Narayanan2021}, which is to leverage its advantage of having two separate decoding passes with different right-contexts. To reduce the latency of showing initial ASR outputs to the users, the first-pass decoding (1st-pass) of the cascaded encoders has zero right-context, while the second-pass decoding (2nd-pass) has a 0.9-second right-context that can improve the quality of the final ASR outputs. Accurate language identification can be achieved by leveraging the 0.9-second right-context, which does not increase the latency of the 2nd-pass. The LID predictor is frame-synchronous and streaming, and the predictions can therefore be included in the input features of either the 2nd-pass decoder or the right-context encoder of the ASR. Experimental results showed that our proposed methods achieved not only accurate LID predictions but reduced WERs as well.

In the rest paper, Sec.~\ref{sec:review} reviews RNN-T with cascaded encoders. Sec.~\ref{sec:methods} presents our proposed methods. Secs.~\ref{sec:exp9} and \ref{sec:exp84} are the experimental setup and results. We conclude in Sec.~\ref{sec:conclusions}. 



\section{ RNN-T with Cascaded Encoders}
\label{sec:review}

First, we will describe the baseline RNN-T cascaded encoder system, which we build our LID prediction on top of.
Given an input utterance $\mathbf{x}_{1:T}$ with $T$ frames\footnote{The term ``frame'' is used in this paper to stand for a decoding time step, which consumes 6 acoustic feature frames \cite{He2019}.} 
and an output sequence $\mathbf{y}=y_{1:U}$ with $U$ subword tokens, RNN-T is an E2E ASR that models $P(\mathbf{y}|\mathbf{x}_{1:T})$ with a single model \cite{Graves2013}.
Regarding a set of subword tokens $\mathcal{V}$, each target in the output layer is either a token in $\mathcal{V}$ or $\varnothing$, where $\varnothing$ is the blank symbol.
Once a non-blank token $y_u$ is emitted, the prediction network of the decoder, usually a long short-term memory (LSTM) model \cite{Graves2013}, is forwarded to derive a text representation based on $y_{1:u}$, which is fused with the acoustic representation derived from the encoder for current frame $t$, and the resulted vector is fed into the output layer to generate the next output symbol. Since one frame would be consumed if $\varnothing$ is generated, the RNN-T decoder is \textit{frame-synchronous}.
In the training, let $\hat{\mathbf{y}}=\hat{y}_{1:T+U}$ be an alignment of $\mathbf{y}$ that can be converted into $\mathbf{y}$ by removing all occurrences of $\varnothing$, $\mathcal{A}(\mathbf{x}_{1:T},\mathbf{y})$ be the reference lattice including all possible alignments between $\mathbf{y}$ and $\mathbf{x}_{1:T}$.
The RNN-T loss  is $\mathcal{L}^\text{rnnt}=\ln P(\mathbf{y}|\mathbf{x}_{1:T})$, where $P(\mathbf{y}|\mathbf{x}_{1:T})$ is
\begin{align*}
P(\mathbf{y}|\mathbf{x}_{1:T})=\sum\nolimits_{\hat{\mathbf{y}}\in\mathcal{A}(\mathbf{x}_{1:T},\mathbf{y})}\prod^{T+U}\nolimits_{i=1}P(\hat{y}_i|\mathbf{x}_{1:t_i},y_{1:u_i}),
\end{align*}
and $t_i$ and $u_i$ are the values of $t$ and $u$ relevant to $\hat{y}_i$ in $\hat{\mathbf{y}}$. 

Due to production needs, we consider only streaming RNN-T, which continuously generates output texts while receiving the audio stream. That is, if a fixed-length right-context with $r\geqslant0$ frames is allowed, only a partial utterance $\mathbf{x}_{1:t+r}$ can be used when calculating $\mathbf{h}^{\text{enc}}_{t}$. Normally, if $r$ increases, the WER reduces; if $r$ decreases, the latency of emitting a new word reduces. To trade-off between WER and latency, \cite{Narayanan2021} proposed an improved RNN-T structure with two cascaded encoders for two separate decoding passes. In the 1st-pass, the causal encoder output $\mathbf{h}^{\text{enc1}}_{t}$ is derived based on $\mathbf{x}_{1:t}$ with zero latency from the right-context. In the 2nd-pass, a right-context encoder is used, whose output $\mathbf{h}^{\text{enc2}}_{t}$ is generated based on $\mathbf{h}^{\text{enc1}}_{1:t+r}$ (and thus $\mathbf{x}_{1:t+r}$). The 2nd-pass decoder achieves lower WERs and higher latency than the 1st-pass decoder with the $r$ frames of extra information. Different from \cite{Narayanan2021} that uses the same decoder for both decodings passes, the two decoders are separate in this paper, which makes the training loss
$\mathcal{L}^\text{casc}$ as
\begin{align}
\label{eq:eq2}
\mathcal{L}^\text{casc}= \lambda \mathcal{L}^\text{1st}+(1-\lambda)\mathcal{L}^\text{2nd},
\end{align}
where $\mathcal{L}^\text{1st}$ and $\mathcal{L}^\text{2nd}$ are the RNN-T loss for 1st- and 2nd-pass, and $\{\lambda, 1-\lambda\}$ are the weights on these two losses respectively.
\section{Joint Language Identification and ASR}
\label{sec:methods}
A frame-synchronous LID predictor is presented in this section which has two benefits. First, it can provide streaming LID predictions at every frame, which can easily be used by the encoder and frame-synchronous decoder of the streaming RNN-T model. Second, the long right-context of the 2nd-pass decoding of cascaded encoders is suitable for predicting LIDs, since accurate language identification often requires a few seconds of audio features that are difficult to do with the conventional low-latency streaming RNN-T. The cascaded encoder structure also allows using complementary features derived from different layers of the causal and right-context encoders to form the inputs of the LID predictor. 

\subsection{Joining a LID predictor with cascaded encoders}

\begin{figure}
     \centering
     \begin{subfigure}[b]{0.75\linewidth}
         \centering
         \includegraphics[width=\linewidth]{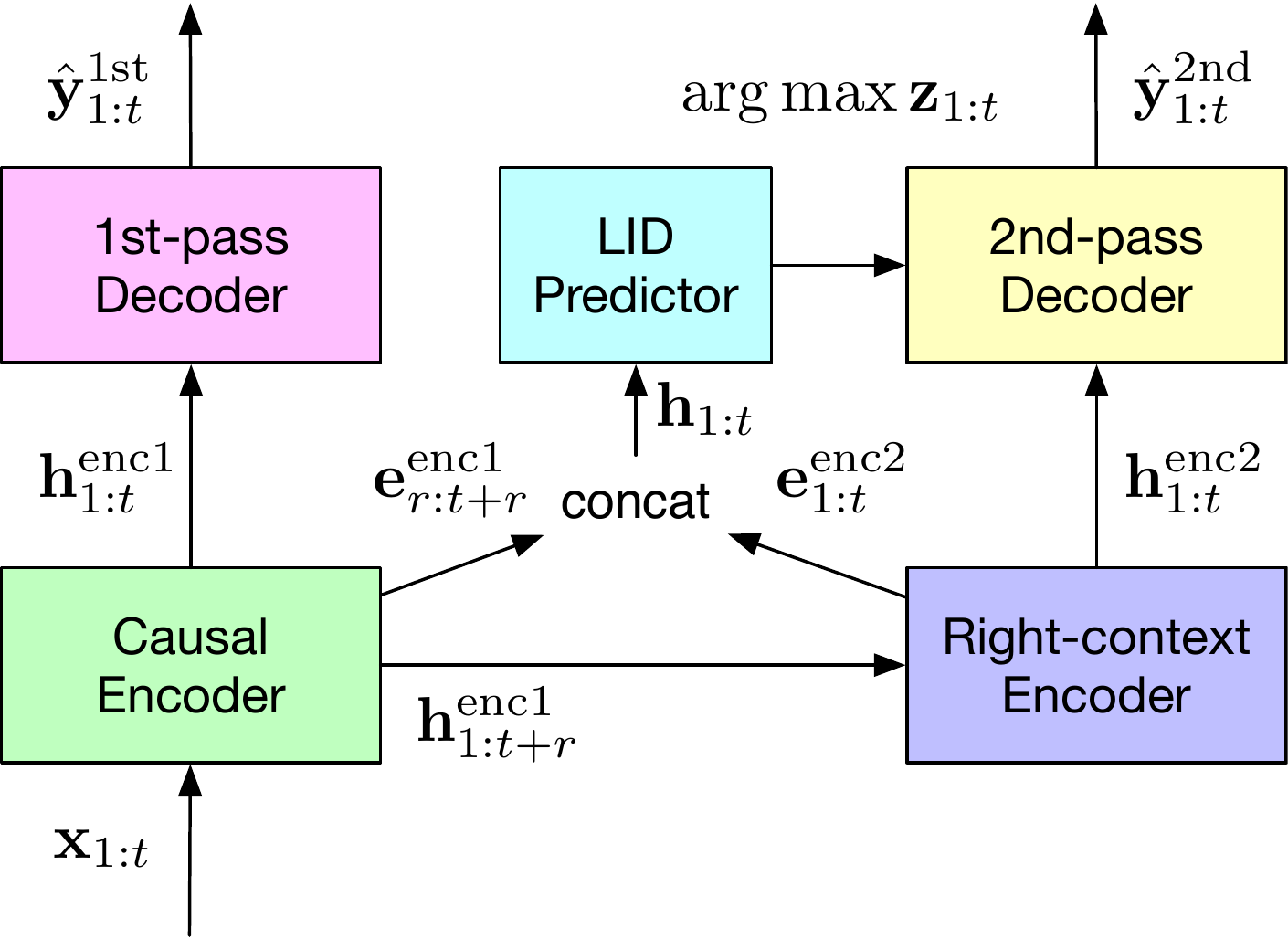}
         \caption{Predicted LIDs fed into the 2nd-pass decoder.}
         \label{fig:struct1}
     \end{subfigure}
     \hfill
     \begin{subfigure}[b]{0.75\linewidth}
         \centering
         \includegraphics[width=\linewidth]{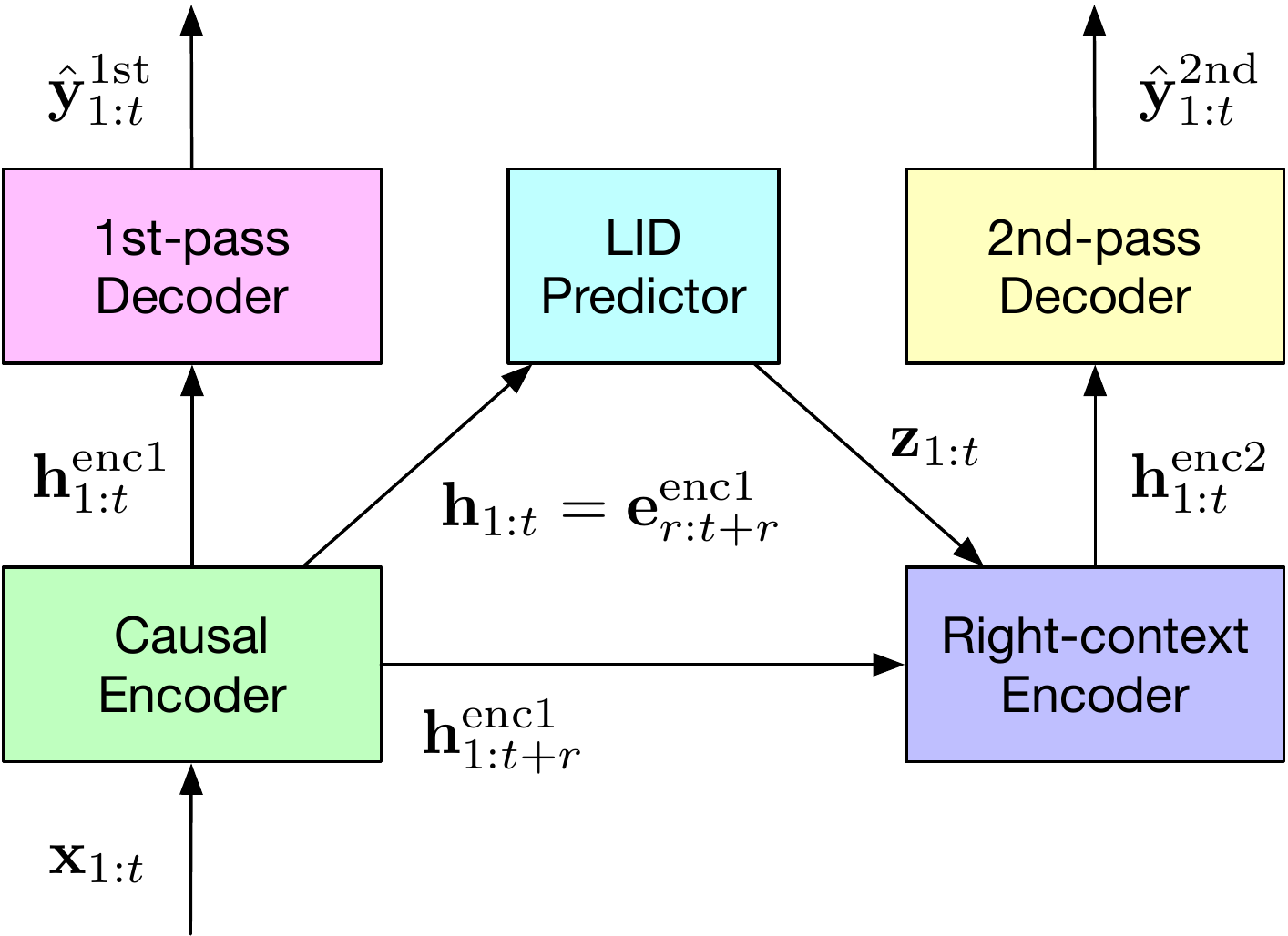}
         \caption{Predicted LIDs fed into the right-context encoder.}
         \label{fig:struct2}
     \end{subfigure}
     \caption{Two ways to incorporate the frame-synchronous LID predictor into RNN-T with cascaded encoders.}
     \vspace{-3mm}
\end{figure}

As discussed before, one of the goals of this paper is to use predicted LIDs to reduce 2nd-pass WERs. As shown in Figs.~\ref{fig:struct1} and \ref{fig:struct2}, the 2nd-pass decoder and right-context encoder are both possible components to incorporate the predicted LIDs, which are both studied and compared later in Secs.~\ref{sec:exp9} and \ref{sec:exp84}.

In the first structure shown in Fig.~\ref{fig:struct1}, the predicted LID distribution sequence $\mathbf{z}_{1:t}$ is converted to a one-hot vector sequence $\arg\max\mathbf{z}_{1:t}$, which is then concatenated with $\mathbf{h}^{\text{enc2}}_{1:t}$ and used as the input features to the 2nd-pass decoder. The input sequence of the LID predictor, $\mathbf{h}_{1:t}$, can thus be obtained by concatenating features derived from both the causal and right-context encoders, which are denoted as $\mathbf{e}^{\text{enc1}}_{1:t+r}$ and $\mathbf{e}^{\text{enc2}}_{1:t}$:
\begin{itemize}
    \item $\mathbf{e}^{\text{enc1}}_{1:t+r}$ is extracted from an earlier causal encoder layer, which is closer to the RNN-T input layer and contains more low-level acoustic information;  
    \item $\mathbf{e}^{\text{enc2}}_{1:t}$ is derived from a final right-context encoder layer, which has more high-level linguistic information as it is closer to the 2nd-pass decoder. 
\end{itemize}
It is known that acoustic and linguistic information can be combined to improve LID prediction \cite{Wang2019,Chandak2020,Punjabi2021}. The concatenation of $\mathbf{e}^{\text{enc1}}_{r:t+r}$ and $\mathbf{e}^{\text{enc2}}_{1:t}$ allows the LID predictor to leverage such complementary information easily. 

Alternatively, as shown in Fig.~\ref{fig:struct2}, LID predictions are fed into the right-context encoder. That is,  $\mathbf{z}_{1:t+r}$ and $\mathbf{h}^{\text{enc1}}_{1:t+r}$ are concatenated to form the input features of the right-context encoder. The input sequence to the LID predictor is $\mathbf{h}_{1:t}=\mathbf{e}^{\text{enc1}}_{r:t+r}$. In addition to the advantage that both 2nd-pass decoder and right-context encoder can benefit from having the predicted LID information, there also exist two disadvantages of this structure. First, only causal encoder features are used for LID prediction, which reduces the diversity of the inputs to the predictor. Second, due to the design of the attention matrices of the streaming Conformer encoder, the right-context allowed for the first encoder layer is often smaller than that allowed for the decoder (\textit{e.g.} 3 frames vs. 15 frames in \cite{He2019,Narayanan2021} and this paper), which further limits the performance of LID predictor.

At last, for both structures, the final training loss $\mathcal{L}^\text{asr+lid}$ is obtained by extending $\mathcal{L}^{\text{casc}}$ to have an additional term for training the LID predictor as a classifier, which is
\begin{align}
\label{eq:eq3}
\mathcal{L}^\text{asr+lid}=\mathcal{L}^\text{casc}+\alpha\,\mathcal{L}^{\text{lid}},
\end{align}
where $\alpha$ is a scalar weight. $\mathcal{L}^{\text{lid}}$ is the cross-entropy loss with 
$\mathcal{L}^{\text{lid}}=\bm{l}_t\ln(\bm{l}_t/\mathbf{z}_{t})$, where $\bm{l}_t$ is the one hot LID label  of $t$.



\subsection{Statistics pooling for streaming LID predictions}


LID prediction is more accurate when using longer audio features \cite{Pesan2016,Tjandra2021}. To facilitate the use of all past frames but still make per-frame predictions, non-parametric statistics pooling \cite{Snyder2018} is used. This pooling converts $\mathbf{h}_{1:t}$ into the concatenation of its mean and standard deviation, $\bm{\mu}_{t}$ and $\bm{\sigma}_{t}$. $\bm{\mu}_{t}$ and $\bm{\sigma}_{t}$ can be computed in a streaming way:
\begin{align}
    \label{eq:eq4}
    \bm{\mu}_{t}&=\theta(\mathbf{h}_{1:t})/t\\
    \label{eq:eq5}
    \bm{\sigma}^2_{t}&=\left(\theta(\mathbf{h}^2_{1:t})-2\bm{\mu}_{t}\theta(\mathbf{h}_{1:t})+t\,\bm{\mu}^2_{t}\right)/t,
\end{align}
where $\theta(\mathbf{h}_{1:t})=\sum^{t}\nolimits_{t'=1}\mathbf{h}_{t'}$  and $\theta(\mathbf{h}^2_{1:t})=\sum^{t}\nolimits_{t'=1}\mathbf{h}^2_{t'}$. For $\bm{\mu}_{t+1}$ and $\bm{\sigma}_{t+1}$, both $\theta(\mathbf{h}_{1:t+1})$ and $\theta(\mathbf{h}^2_{1:t+1})$ are obtained easily by adding $\mathbf{h}_{t+1}$ and $\mathbf{h}^2_{t+1}$ to $\theta(\mathbf{h}_{1:t})$, $\theta(\mathbf{h}^2_{1:t})$. 
After converting $\mathbf{h}_{1:t}$ into a concatenated vector $[\bm{\mu}_{t}; \bm{\sigma}_t]$ with the statistics pooling,  another two fully connected (FC) layers followed by a softmax output layer are used to transform $[\bm{\mu}_{t}; \bm{\sigma}_t]$ into the LID distribution $\mathbf{z}_t$. As a result, our frame-synchronous LID predictor is efficient for handling streaming data that only requires a small amount computational cost during test.

\section{Experimental Setup}
\label{sec:exp9}

\subsection{Data}
Experiments were conducted on a dataset with 9 language locales collected from Google’s Voice Search traffic, which are: US English (en-US), UK English (en-GB), French (fr-FR), Italian (it-IT), Germany (de-DE), US Spanish (es-US), ES Spanish (es-ES), Taiwan Chinese (zh-TW) and Japanese (ja-JP). The training data is mixed with all language locales. There are in total 214.2M utterances corresponding to 142.3K hours of speech data. en-US and en-GB take about 25\% and 5\% of the training data, while each of the rest 7 locales takes about 10\% of the data. The SpecAugment method \cite{Park2019} is used to improve ASR robustness against noisy conditions. The testing utterances are also sampled from Google’s Voice Search traffic with a maximum duration constraint of 5.5-second long for each utterance. The test sets have no overlapping from the training set for evaluation purposes. All data are anonymised and hand-transcribed, and abide by Google AI Principles \cite{googleaiprinciples}. In addition, all data is normalised to have zero-mean and unit-variance based on the statistics collected from the whole training set. 

\subsection{Model}
\label{ssec:setup2}
The acoustic features are 80-dimensional (-dim) log-Mel filter bank features computed using a 32ms frame length and a 10ms shift. Acoustic feature vectors from 3 contiguous frames are stacked and subsampled to form a 240-dim input vector with a 30ms frame rate, which is then transformed using a linear projection to 512-dim and added with positional embeddings. 

Regarding the causal encoder, twelve Conformer encoder blocks with 8-head self-attention and a convolution kernel size of 15 are used to further transform the stacked features. A concatenation operation is performed after the 3rd block to achieve a time reduction rate of 2, and the resulted 1024-dim vectors are transformed by the 4th Conformer block and then projected back to 512-dim using another linear transform. Afterward comes with another 8 Conformer blocks followed by a final linear normalisation layer. These layers combined together form the causal-encoder with 110M parameters. The right context of every causal-encoder layer is 0 frame.

For the right-context encoder, a 512-dim linear projection is first used to transform its input features, followed by five 512-dim Conformer blocks and a final linear normalization layer, which altogether results in 50M parameters. The total right context of the right-context-encoder is 15 frames, which results in a 0.9-second latency of the second-pass decoder. Such 15-frame right-context is evenly distributed to the five Conformer blocks, with each of them having a 3-frame right context. 

There are in total 16,384 word pieces used as the subword tokens for the 9 language locales combined and are modelled with a single softmax output layer following \cite{Li2021}. The two separate decoders use the same structure with 33M parameters each, which consists of two layers of 2,048-dim LSTM with a 640-dim linear projection in the prediction network. Our LID predictor uses two 512-dim FC layers that result in 0.7M--1M parameters depending on the size of its input feature sequence, which takes less than 0.5\% of the total model parameters. 



All models are trained in Tensorflow with the Lingvo toolkit  on Google’s TPU V3 with a global batch size of 4,096 utterances. Models are optimized using synchronized stochastic gradient descent based on the Adam optimiser with $\beta_1=0.9$ and $\beta_2=0.999$.  $\lambda$ and $\alpha$ in Eqns.~\eqref{eq:eq2} and \eqref{eq:eq3} are set to 0.5 and 0.05. The models are tested without any external LMs.


  

\section{Experimental Results}
\label{sec:exp84}

\setcounter{table}{3}
\begin{table*}[!htbp]
    \caption{800K-step LID prediction \%accuracy breakdown to individual languages/locales. ``en-X'' is the cluster of ``en-GB'' \& ``en-US'', and ``es-X'' is the cluster of ``es-ES'' \& ``es-US''. The systems used are S20 with or without statistics pooling. ``Frame'' is the frame index $t$ of $\mathbf{e}^{\text{enc2}}_{1:t}$ in Fig.~\ref{fig:struct1}, and ``avg.'' refers to the average of all possible $t$.}
    \vspace{-3mm}
    \centering
    \begin{tabular}{cccccccccccccc}
        \toprule
        Pooling & Frame & de-DE & en-GB & en-US & en-X & es-ES & es-US & es-X & fr-FR & it-IT & ja-JP & zh-TW & Mean \\
        \midrule
        w/ & avg. &  95.5 & 82.0 & 96.8 & -- & 87.3 & 89.7 & -- & 95.3 & 94.6 & 97.4 & 97.2 & \textbf{92.9} \\
        w/o & avg. & 92.3 & 75.2 & 95.5 & --  & 81.3 & 84.5 & -- & 91.7 & 90.5 & 95.3 & 95.1 & 89.0  \\
        \midrule
         & avg. & 95.5 & -- & -- & 97.5 & -- & -- & 96.0 & 95.3 & 94.6 & 97.4 & 97.2 &  96.2 \\
        & 0 & 86.2 & -- & -- & 92.7 & -- & -- & 86.2 & 85.6 & 80.8 & 92.6 & 92.4 & 88.1 \\
        w/ & 15 & 93.7 & -- & -- & 96.3 & -- & -- & 94.0 & 93.8 & 91.7 & 96.9 & 96.6 & 94.7 \\
        & 30 & 97.1 & -- & -- & 98.2 & -- & -- & 97.3 & 97.1 & 96.2 & 98.8 & 98.5 & 97.6 \\
        & $\infty$ & 98.5 & -- & -- & 99.1 & -- & -- & 98.7 & 98.5 & 98.3 & 99.2 & 99.0 & 98.8 \\
        \bottomrule
    \end{tabular}
    \label{tab:lidresults}
    \vspace{-3mm}
\end{table*}
\setcounter{table}{0}
\subsection{Summary of 800K-step ASR results}
\label{ssec:systems}
ASR results of the baseline and proposed cascaded encoder systems are listed in Table~\ref{tab:systems}. S00 is the system without using any LID information. S01--S02 are those with the oracle one-hot LID vectors being used to augment the input features of different model components for both training and test.
S20 and S30 are with our proposed structures shown in Fig.~\ref{fig:struct1} and Fig.~\ref{fig:struct2} respectively. Comparing S01 with S00, including oracle LID information throughout model reduced 1st-pass and 2nd-pass WERs both by a relative of 3\%, which is the maximum WER reduction can be achieved with predicted LIDs. Compared with S01, S20 achieved the same 2nd-pass WER, showing the effectiveness of the proposed structure that feeds the predicted LID into the 2nd-pass decoder. S20 1st-pass WER is similar to S00 and worse than S01, which is expected since the predicted LIDs are not used in the 1st-pass. The estimated 2nd-pass WER gain of our proposed system S20 compared to the baseline S00 is 0.2\% (computed from $\sim$0.4M utterances across the 9 locales), which is \textbf{statistically significant} at the 5\% level. 
Furthermore, since the 2nd-pass WER is the final WER of the system, the Table shows that we can achieve the first goal of this paper, namely that the same final ASR accuracy can be obtained using predicted (S20) and oracle LIDs (S01). 

The WER reduction over S00 obtained by S30 is only 33\% of that achieved by S20. When reducing the right-context from 15 to 3 to meet the 3-frame right-context of the first Conformer block of the right-context encoder, the resulting S31 system WERs were slightly better than S00 and received an improvement in ASR accuracy with the predicted LIDs. These reveal that the structure in Fig.~\ref{fig:struct1} is better than that in Fig.~\ref{fig:struct2}. Furthermore, S20 can be viewed as replacing the oracle LIDs in S02 with the predicted ones. 
Since S20 had lower 2nd-pass WER than S02, not only including the predicted LIDs in the 2nd-pass decoder, but also having a shortcut for gradient back-propagation between the 2nd-pass decoder and causal encoder helped to reduce 2nd-pass WER. 
In the next section, we will validate this finding by analysing S20. 



\begin{table}[h!]
    \caption{800K-step ASR results of key systems, where ``1st'' and ``2nd'' are 1st- and 2nd-pass decoding results in \%WER. }
    \vspace{-3mm}
    \centering
    \begin{tabular}{llcc}
        \toprule
        {ID} & {System} & {1st} & {2nd}  \\
        \midrule
        S00 & no LID & 9.1 & 8.0 \\
        S01 & oracle LID to causal encoder & \textbf{8.9} & \textbf{7.8} \\
        S02 & oracle LID to 2nd-pass decoder & 9.1 & 7.9 \\
        \midrule
        S20 & Fig.~\ref{fig:struct1} system with $r$=15  & {9.1} & \textbf{7.8} \\
        S30 & Fig.~\ref{fig:struct2} system with $r$=15 & 9.1 & 7.9 \\
        S31 & Fig.~\ref{fig:struct2} system with $r$=3 & 9.1 & 7.9 \\
        \bottomrule
    \end{tabular}
    \label{tab:systems}
    \vspace{-3mm}
\end{table}

\subsection{Input and output settings for the LID predictor}
\label{ssec:lidsetting}
In this section, S20 is analysed to understand the impact of different input and output settings. 
All results in Table~\ref{tab:lidinput} and \ref{tab:lidoutput} are produced by systems trained for 200K steps, instead of the final 800K-step systems used in Table~\ref{tab:systems}.
Table~\ref{tab:lidinput} shows the results of deriving $\mathbf{e}^{\text{enc1}}_{1:t+r}$ and $\mathbf{e}^{\text{enc2}}_{1:t}$ from different encoder layers to form $\mathbf{h}_{1:t}$ in Fig.~\ref{fig:struct1} systems. Deriving $\mathbf{e}^{\text{enc1}}_{1:t+r}$ from the 4th Conformer block resulted in lower 1st-pass WERs due to the regularisation effect. By combining that $\mathbf{e}^{\text{enc1}}_{1:t+r}$ with the $\mathbf{e}^{\text{enc2}}_{1:t}$ derived from the 5th Conformer block of the right-context encoder, higher LID prediction accuracy can be achieved that leverages the complementary acoustic and linguistic information. This setting (last row of Table~\ref{tab:lidinput}) is used in S20.

\begin{table}[!htbp]
    \caption{200K-step Fig.~\ref{fig:struct1} system results using different inputs for the LID predictor, where b$_n$ refers to outputs from the $n$-th Conformer block.
    ``1st'' and ``2nd'' are the 1st-pass and 2nd-pass decoding results in \%WER; ``9-cls'' is the LID \%accuracy averaged over the 9 language locales and across time.}
    \vspace{-3mm}
    \centering
    \begin{tabular}{llcccc}
        \toprule
        $\mathbf{e}^{\text{enc1}}_{1:t+r}$  & $\mathbf{e}^{\text{enc2}}_{1:t}$ & {1st} & {2nd} & {9-cls}  \\
        \midrule
         b$_4$ & -- & 9.9 & 8.5 & 89.8 \\
         -- & b$_5$ & 10.0 & 8.5 & 90.6 \\
         b$_4$ & b$_5$ & \textbf{9.9} & \textbf{8.5} & \textbf{91.5}\\
        \bottomrule
    \end{tabular}
    \label{tab:lidinput}
\end{table}


Next, different ways of feeding the predicted LID distribution $\mathbf{z}_{1:t}$ into the 2nd-pass decoder are compared by training different Fig.~\ref{fig:struct1} systems. Compared with feeding $\mathbf{z}_{1:t}$ directly, feeding $\arg\max\mathbf{z}_{1:t}$ as a sequence of 9-dim one-hot vectors achieved lower WERs and higher LID prediction accuracy. When feeding  $\text{sg}(\arg\max\mathbf{z}_{1:t})$, where $\text{sg}(\cdot)$ means the input tensor will have zero gradients, the shortcut created by the LID predictor between the 2nd-pass decoder and causal encoder is removed. Although this improves the 1st-pass WER and LID prediction accuracy, it increases the 2nd-pass WER. This shows that some improvements in the 2nd-pass decoding when using $\arg\max\mathbf{z}_{1:t}$ come from the shortcut. Finally, clustering ``en-GB'' and ``en-US'' into ``en-X'' and clustering ``es-ES'' and ``es-US'' into ``es-X'', the 9 language locales are merged into 7 distinct languages. By using the 7 languages instead of 9 locales in both training and test, the resulted system (last row of Table~\ref{tab:lidoutput}) has worse performance compared to $\arg\max\mathbf{z}_{1:t}$. As a result, $\arg\max\mathbf{z}_{1:t}$ is used for S20 in Table~\ref{tab:systems} to achieve a good trade-off between the 2nd-pass WER and LID prediction accuracy.

\begin{table}[!htbp]
    \caption{200K-step Fig.~\ref{fig:struct1} system results with different ways of using predicted LIDs in the 2nd-pass decoder.
    ``1st'' and ``2nd'' are the 1st-pass and 2nd-pass decoding results in \%WER. ``9-cls'' and ``7-cls'' are the LID prediction results in \%accuracy avg. across time for the 9 language locales and 7 languages.}
    \vspace{-3mm}
    \centering
    \begin{tabular}{ccccc}
        \toprule
        {System} & {1st} & {2nd} & 9-cls & 7-cls \\
        \midrule
        $\mathbf{z}_{1:t}$ & 10.0 & 8.6 & 91.3 & 95.3 \\
        $\arg\max\mathbf{z}_{1:t}$ & 9.9 & \textbf{8.5} & 91.5 & \textbf{95.5} \\
        $\text{sg}(\arg\max\mathbf{z}_{1:t})$ & {9.9} & 8.6 & \textbf{91.6} & 95.4 \\
        $\text{cluster}(\arg\max\mathbf{z}_{1:t})$ & {10.0} & 8.6 & -- & 95.4  \\
        \bottomrule
    \end{tabular}
    \label{tab:lidoutput}
    \vspace{-4mm}
\end{table}

\subsection{Analysis of streaming LID predictions}
\label{ssec:lidresult}
At last, detailed LID prediction results of Fig.~\ref{fig:struct1} structured systems are given in Table~\ref{tab:lidresults}. From the results, S20 achieved a 92.9\% LID classification accuracy averaged across time, which outperformed S20 without statistics pooling (2nd row in Table~\ref{tab:lidresults}). By further evaluating the clustered 7 languages, the averaged LID prediction accuracy of S20 reached 96.2\%. Evaluating LID performance at different frames in the LID prediction sequence, the LID accuracy increases when more frames are available in a streaming utterance, and the final LID accuracy based on complete utterances is 98.8\%. In particular, from 0th frame to 15th frame, ``de-DE'', ``es-X'', ``fr-FR'', and ``it-IT'' received more obvious increases in accuracy than ``en-X'', ``ja-JP'' and ``zh-TW'', which is perhaps due to their similarity to ``en-X'' that can require longer feature sequences to be well discriminated. These results show that by adding an efficient LID predictor, RNN-T with cascaded encoders can produce LID accurately in a streaming way, which fulfills our second goal. 

\section{Conclusions}
\label{sec:conclusions}
We propose to build a truly multilingual E2E ASR system that combines ASR with LID prediction based on the structure of RNN-T with cascaded encoders. By leveraging on the long right-context in the second-pass decoding that is useful for LID prediction, a lightweight predictor is added to the ASR structure to achieve accurate frame-synchronous LID prediction in a streaming way. Experiments on a large-scale multilingual voice search setup with 9 language locales showed, by only increasing 0.5\% of the number of model parameters and little test-time cost, the LID predictor achieved an average of 96.2\% accuracy across time when evaluated on 7 languages clustered from the 9 locales. By feeding the predicted LID into the second-pass decoder, our multilingual ASR that does not require knowing LIDs can achieve the same final WER as the system that includes the oracle LID as extra input. 
It is possible to apply our method to handle code-switch with mixed language data. 


\vfill\pagebreak
\section{References}
\label{sec:refs}
\renewcommand{\section}[2]{}
\renewcommand*{\bibfont}{\footnotesize}
\setlength{\bibsep}{3pt plus 0.3ex}
\bibliographystyle{IEEEtran}
\bibliography{mybib}

\end{document}